\documentstyle[11pt,seceq]{article}

 \newcommand{\be}{\begin{equation}}
 \newcommand{\ee}{\end{equation}}
 \newcommand{\lb}{\label}
 \newcommand{\bff}{{\bf f}}
 \newcommand{\bh}{{\bf h}}
 \newcommand{\bk}{{\bf k}}
 \newcommand{\bq}{{\bf q}}
 \newcommand{\bm}{{\bf m}}
 \newcommand{\bx}{{\bf x}}
 \newcommand{\bz}{{\bf z}}
 \newcommand{\hL}{{\hat{{\cal L}}}}
 \newcommand{\bA}{{\bf A}}
 
 \newcommand{\bC}{{\bf C}}
 \newcommand{\bG}{{\bf G}}
 \newcommand{\bH}{{\bf H}}
 
 \newcommand{\bK}{{\bf K}}
 \newcommand{\bL}{{\bf L}}
 
 \newcommand{\bQ}{{\bf Q}}
 \newcommand{\bR}{{\bf R}}

 \newcommand{\bV}{{\bf V}}
 \newcommand{\bW}{{\bf W}}
 
 \newcommand{\bhG}{{\widehat{{\bf G}}}}
 
 \newcommand{\bhU}{{\widehat{{\bf U}}}}
 \newcommand{\bhV}{{\widehat{{\bf V}}}}
 \newcommand{\bhZ}{{\widehat{{\bf Z}}}}
 \newcommand{\bhq}{{\widehat{{\bf q}}}}
 
 \newcommand{\bdot}{{\mbox{\boldmath $\cdot$}}}
 \newcommand{\grad}{{\mbox{\boldmath $\nabla$}}}
 \newcommand{\bzed}{{\mbox{\boldmath $0$}}}

 \newcommand{\balpha}{{\mbox{\boldmath $\alpha$}}}
 \newcommand{\bpsi}{{\mbox{\boldmath $\psi$}}}
 \newcommand{\bmu}{{\mbox{\boldmath $\mu$}}}
 \newcommand{\bhpsi}{\widehat{{\mbox{\boldmath $\psi$}}}}
 \newcommand{\bhPsi}{\widehat{{\mbox{\boldmath $\Psi$}}}}
 \newcommand{\hpsi}{\widehat{\psi}}
 \newcommand{\AAA}{{\cal A}}
 \newcommand{\PPP}{{\cal P}}
 \newcommand{\QQQ}{{\cal Q}}
 \newcommand{\bGamma}{{\mbox{\boldmath $\Gamma$}}}
 \newcommand{\rT}{{\rm T}}

\begin{document}
 \title{Linear Stochastic Models of Nonlinear Dynamical Systems}
 \author{Gregory L. Eyink \\
         {\em Department of Mathematics}\\
         {\em University of Arizona}\\
         {\em Tucson, AZ, 85721}}

 \date{ }
 \maketitle
 \begin{abstract}
  We investigate in this work the validity of linear stochastic models for
nonlinear dynamical systems.
  We exploit as our basic tool a previously proposed Rayleigh-Ritz
approximation for the {\em effective action}
  of nonlinear dynamical systems started from random initial conditions. The
present paper discusses only the case
  where the PDF-{\em Ansatz} employed in the variational calculation is
``Markovian'', i.e. is determined
  completely by the {\em present} values of the moment-averages. In this case
we show that the Rayleigh-Ritz
  effective action of the complete set of moment-functions that are employed in
the closure has a quadratic part
  which is always formally an {\em Onsager-Machlup action}. Thus, subject to
satisfaction of the requisite
  realizability conditions on the noise covariance, a linear Langevin model
will exist which reproduces
  exactly the joint 2-time correlations of the moment-functions. We compare our
method with the closely related
  formalism of {\em principal oscillation patterns} (POP), which, in the
approach of C. Penland, is a method
  to derive such a linear Langevin model empirically from time-series data for
the moment-functions. The predictive
  capability of the POP analysis, compared with the Rayleigh-Ritz result, is
limited to the regime of small
  fluctuations around the most probable future pattern. Finally, we shall
discuss a {\em thermodynamics of
  statistical moments} which should hold for all dynamical systems with stable
invariant probability measures
  and which follows within the Rayleigh-Ritz formalism.
 \end{abstract}

\newpage

\section{Introduction}

We consider nonlinear dynamical systems governed by (possibly nonautonomous)
differential equations:
\be \dot{\bx}=\bhU(\bx,t). \lb{a} \ee
We may take $\bx=(x_1,...,x_p)^\top$ to have any number of components, possibly
infinitely many, formally
including infinite-dimensional dynamical systems governed by partial
differential equations, etc.  In many
contexts the dynamics of a selected set of variables
$\bhpsi(\bx)=(\hpsi_1(\bx),...,\hpsi_n(\bx))^\top$
is of interest. Of course, by the chain rule
\be \partial_t\bhpsi = (\bhU\cdot\grad_\bx)\bhpsi := \bhV. \lb{b} \ee
When the dynamics is nonlinear, the righthand side $\bhV$ of equation (\ref{b})
cannot generally be expressed
in terms of the functions $\bhpsi$ themselves.  For example, when $\bhU(\bx)$
and $\bhpsi(\bx)$ are polynomial
functions of $\bx$, the righthand side consists of higher-degree polynomials.
This is a manifestation of
the {\em closure problem} of nonlinear dynamical systems.

If one considers the initial-value problem with random initial data, a common
strategem
to obtain mean values $\bm(t):=\langle \bhpsi\rangle_t$ is to make a {\em
moment closure} approximation
$\langle(\bhU\cdot\grad_\bx)\bhpsi\rangle_t\approx \bV(\bm,t)$ for some
function $\bV$ of the selected moments,
so that a closed equation
\be \dot{\bm}(t) = \bV(\bm,t) \lb{c} \ee
is obtained. Likewise, if one is interested in {\em fluctuations} of the
variables $\bhpsi(t)$, then one
can make an approximation that
\be \dot{\bhpsi}(t) \approx \bV(\bhpsi,t) + \bhq(t) \lb{d} \ee
where $\bhq(t)$ is a {\em random force} of known statistics, which is supposed
to represent the effects of
neglected variables beyond the subset $\bhpsi$ retained. Such a model---if it
is valid---will clearly give
important information about predictability of the variables $\bhpsi(t)$. For
example, conditional probabilities
$P(\bpsi,t|\bpsi_0,t_0)$ are implied, which express precisely the limits on
predicting the moment-variables
at a future time $t$ given their values at the present time $t_0$. One is thus
interested to know the possibilities
and limitations of such a stochastic modelization.

In previous work \cite{Ey96,Ey98} we have studied fluctuations of nonlinear
dynamics by an {\em action principle}. Such
an approach to fluctuation theory goes back to the work of Onsager and Machlup
\cite{OM}. They showed that a {\em linear
Langevin dynamics}
\be \dot{\bhpsi}(t) = \bA(t)\bhpsi + \bhq(t), \lb{e} \ee
in which $\bhq(t)$ is a Gaussian random force with zero mean and covariance
\be \langle\bhq(t)\bhq^\top(t')\rangle = \bQ(t)\delta(t-t'), \lb{ee} \ee
can always be completely and equivalently reformulated in terms of an action
functional $\Gamma[\bpsi]$:
\be \Gamma[\bpsi]={{1}\over{4}}\int_{t_0}^\infty
dt\,\,(\dot{\bpsi}-\bA\bpsi)^\top
                                 \bQ^{-1}(\dot{\bpsi}-\bA\bpsi). \lb{eee} \ee
The interpretation of this functional is as a ``fluctuation potential'' for
time-histories. That is, the probability
that a particular fluctuation value $\bpsi(t)$ occur for the random variable
$\bhpsi(t)$ is given in terms of the
Onsager-Machlup action by the exponential formula
\be {\rm Prob}\left(\bhpsi(t)\approx \bpsi(t):-\infty<t<+\infty\right)\sim
e^{-\Gamma[\bpsi]}. \lb{f} \ee
This gives the most direct probabilistic significance of the Onsager-Machlup
action. The fact that the action is
a quadratic functional of $\bpsi$ is consistent with the fact that the solution
of the linear Langevin equation
is a normal random variable, with a Gaussian probability distribution.

Although the Onsager-Machlup theory as originally developed was restricted to
linear Langevin dynamics, the action
method is completely general: For any statistical dynamical system and for any
selected subset of random variables
$\widehat{\bpsi}$, an {\em effective action} $\Gamma[\bpsi]$ can be introduced
which plays the same role as the
Onsager-Machlup action does for linear Langevin dynamics. It has too an
interpretation as a fluctuation potential
\footnote{The additional factor of $N$ in the exponent in (\ref{34c}) is
discussed more below. See
Eqs. (\ref{34b})-(\ref{34e}).} for the empirical average over $N$ independent
samples \cite{Cram}, i.e.
\be {\rm Prob}\left({{1}\over{N}}\sum_{n=1}^N \bhpsi^{(n)}(t)\approx
\bpsi(t):-\infty<t<+\infty\right) \sim
                  \exp\left(-N\cdot \Gamma[\bpsi]\right). \lb{34c} \ee
The effective action is also a generating functional for all (irreducible)
multi-time correlations of the variables
$\bhpsi(t)$, of arbitrary order, and thus completely characterizes the
distribution of those variables. To be precise, if
the effective action is expanded into a functional power series in
$\delta\bpsi(t):=\bpsi(t)-\langle\bhpsi(t)\rangle$, as
\be \Gamma_*[\bpsi]= \sum_{k=2} {{1}\over{k!}}\int dt_1\cdots\int dt_k
\,\,\Gamma^{(k)}_{i_1\cdots i_k}(t_1,...,t_k)\,\,
                     \delta\psi_{i_1}(t_1)\cdots \delta\psi_{i_k}(t_k), \lb{g}
\ee
then the coefficients are just the irreducible multi-time correlators
\cite{IZ}. The correlators with $k\geq 3$ would all be
zero for a Gaussian process. We shall not review these subjects further here,
since they have been thoroughly discussed
elsewhere \cite{Ey96,Ey98}.

In our earlier works, we developed a {\em Rayleigh-Ritz} approximation method
by which the effective actions $\Gamma[\bz]$
of {\em any} set of random variables $\bhZ$ may be calculated within a
PDF-based moment-closure scheme. In particular, an
approximate effective action may be obtained for $\bhpsi$, the moment-variables
retained in the closure. We shall show
here that such a Rayleigh-Ritz approximate effective action $\Gamma_*[\bpsi]$
of the moment-variables themselves is not only
a formal generalization of the Onsager-Machlup action, but is actually much
more closely related. In fact, we shall show
that the leading quadratic term in the Taylor expansion (\ref{g}) of
$\Gamma_*[\bpsi]$ is always precisely of the
Onsager-Machlup form, when the PDF-{\em Ansatz} employed in the Rayleigh-Ritz
calculation is ``Markovian''. By the
latter specification we denote PDF-{\em Ans\"{a}tze} which are completely
determined by the {\em present} values
which they assign to averages of the moment functions. Our result means that,
{\em for such a ``Markovian'' PDF}-Ansatz,
{\em there is always formally a linear Langevin dynamics such as (\ref{e})
which gives predictions for the 2-time
correlations} $\langle\delta\bhpsi(t)\delta\bhpsi^\top(t')\rangle$ {\em that
are the same as those given by the
Rayleigh-Ritz effective action}. In general, however, for higher-order
correlations, the linear Langevin model
and the Rayleigh-Ritz effective action will not yield the same predictions.

It is our purpose here to present the derivation of the linear Langevin model
via the effective action method and to
discuss its physical interpretation and limits of applicability. The effective
action provides a framework to derive not
only the linear theory but also the higher-order statistics (higher order in
terms of the size of the fluctuations or
the order of the correlator). It thus provides a means to assess the size of
the corrections to the linear description.
On the other hand, the linear Langevin model gives always the leading-order
contribution to the effective action and,
therefore, many of the important features of the full Rayleigh-Ritz
approximation are essentially entirely determined
by the linear equation. We shall in addition give a more intuitive derivation
of the Langevin model within moment-closure
methodology, but not using a systematic or formal scheme. While such a
derivation provides no possibility to assess
limitations of the linear description, nevertheless it provides insight into
the physical assumptions involved.
We shall also compare our method with the {\em principal oscillation pattern
(POP)} analysis, which is a well-known
method to extract linear stochastic models empirically from time series data
\cite{Pen}. Finally, we shall conclude
with some general discussion on the thermodynamics of moment-averages for
dynamical systems with stable statistics.
In particular, we discuss the law of entropy increase and
fluctuation-dissipation relations at the linear level.

\section{Rayleigh-Ritz Effective Action of Moment Variables}

\noindent {\em (2.1) Reprise of the Rayleigh-Ritz Method}

\noindent The Rayleigh-Ritz approximation of the effective action is based upon
a variational formulation of the moment
closure scheme. This is just a variational formulation of the method of
weighted residuals \cite{Finlay} to solve
$\partial_t\PPP= -\grad_\bx\bdot (\bhU \PPP) :=\hL\PPP$, the Liouville equation
for the phase-space distribution $\PPP$.
The basic ingredients of a moment closure are (i) a set of {\em moment
functions} $\bhpsi(\bx,t)=(\hpsi_1(\bx,t),
...,\hpsi_n(\bx,t))$ and (ii) a PDF {\em Ansatz} $\PPP(\bx;\bm,t)$,
conveniently parametrized by
the mean values $m_i=\langle\hpsi_i\rangle,\,\,\,i=1,...,n$ which it assigns to
those functions.
The variable $t$ is included to denote an {\em explicit} time-dependence, i.e.
any time-dependence
other than the implicit one through parameters $\balpha(t),\bm(t)$. The
$\balpha$-type parameters
appear in the variational formulation of the closure, in which one incorporates
all of the moment functions
into a single linear combination,
\be \AAA(\bx;\balpha,t)=\sum_{i=0}^n \alpha_i \hpsi_i(\bx,t). \lb{1} \ee
Note that the constant function $\hpsi_0(\bx,t)\equiv 1$ must be included in
the sum in order
to satisfy the final-time condition $\AAA(\infty)\equiv 1$. The PDF {\em
Ansatz} $\PPP(\bx;\bm,t)$ is subject
to an initial-condition that it match the considered initial distribution
$\PPP_0$ for the problem,
$\PPP(\bx;\bm,t_0)= \PPP_0$, in the weighted-residual sense that the averages
of the $n$ moment-functions $\bhpsi(t)$
must match. Then, it is not hard to show that the {\em moment closure equation}
\be \dot{\bm}(t) = \bV(\bm(t),t) \lb{1a} \ee
where $\bV(\bm,t):=\langle(\partial_t+\hL^\dagger)\bhpsi(t)\rangle_{\bm,t}$, is
the result of varying the
{\em action functional}
\be \Gamma[\AAA,\PPP] = \int_{t_0}^\infty
dt\,\,[\langle\AAA(t),\dot{\PPP}(t)\rangle-\langle\AAA(t),\hL\PPP(t)\rangle ]
     \lb{2} \ee
over the above {\em Ans\"{a}tze} for $\AAA(t),\PPP(t)$, with variational
parameters $\alpha_0(t),\balpha(t),\bm(t)$.
The Euler-Lagrange equations for $\bm(t)$ are just (\ref{1a}) while the
equations for $\alpha_0(t),\balpha(t)$
have the unique solution $\alpha_0(t)\equiv 1,\balpha(t)\equiv\bzed$ subject to
the final-conditions.

The Rayleigh-Ritz approximation to the effective action $\Gamma_*[\bz]$ of a
set of random variables $\bhZ$
is obtained, in general, as the stationary point $\Gamma_*[\bz] = {\rm
st.pt.}_{\AAA,\PPP}\Gamma[\AAA,\PPP]$
varied over $\AAA(t),\PPP(t)$ of the above forms, subject to the additional
constraints of unit overlap
\be \langle\AAA(t),\PPP(t)\rangle=1 \lb{3} \ee
and fixed expectation
\be \langle\AAA(t),\bhZ(t)\PPP(t)\rangle= \bz(t) \lb{4} \ee
for each given history $\bz(t)$, for all times $t$ after the initial time
$t_0$. \footnote{Recall that
$\bhZ(t)$ is an observable in ``Schr\"{o}dinger picture'' and that the only
time-dependence is explicit.}

We show here that the Rayleigh-Ritz approximation $\Gamma_*[\bpsi]$ to the
effective action of the moment-variables
themselves has, in general, a quadratic part which is just an Onsager-Machlup
action, when closure is achieved
within the framework outlined above. Other closure schemes are conceivable
within the Rayleigh-Ritz formalism
and may even better represent the physics in certain situations. A
``Markovian'' approximation has been made above,
in assuming that the PDF {\em Ansatz} $\PPP(\cdot;\bm,t)$ is parameterized by
only the {\em present value} $\bm(t)$
of the $n$ moment-averages. This is by no means necessary. More generally, one
may assume that $\PPP[\cdot;\bm,t]$
is a functional of the entire {\em past mean-history} $\{\bm(s): s<t\}$ of the
$n$ moment-functions. In that case,
the closure equation becomes
\be \dot{\bm}(t) = \bV[t;\bm], \lb{1aa} \ee
in which $\bV[t;\bm]$ is now also a functional over the past mean-history. It
is not hard to show that an {\em exact}
equation always exists of the form (\ref{1aa}) for a suitable choice of the
functional $\bV[t;\bm]$. (E.g. see \cite{Lum},
Appendix A1.) Thus, a closure incorporating such history effects is likely to
be more faithful to the physics, in general.
Our work here does not discuss this more general case, but confines itself to
the ``Markovian'' {\em Ansatz}. Although
this is restrictive, it is nevertheless the case that most practical closures
considered in the literature are of this
type. It is also possible, even within this more restrictive ``Markovian''
framework, to include some history effects.
This may be done, for example, by constructing a closure using not only the $n$
moment-functions $\bhpsi(t)$,
but also the corresponding $n$ {\em velocity-functions} \footnote{This
definition generalizes that in Eq.(\ref{b})
to the case with explicit time-dependence.}
\be \bhV(t):=(\partial_t + \hL^\dagger)\bhpsi(t). \lb{4a} \ee
In this case, the closure equations become, instead of (\ref{1a}),
\begin{eqnarray}
\dot{\bm}(t) & = & \bV(t) \cr
\dot{\bV}(t) & = & \bG(\bm(t),\bV(t),t), \lb{4b}
\end{eqnarray}
where $\bG(\bm,\bV,t):=\langle(\partial_t +
\hL^\dagger)\bhV(t)\rangle_{\bm,\bV,t}$ is the average
{\em acceleration} of the moment-functions within a PDF-{\em Ansatz} depending
jointly upon $\bm,\bV$.
Such schemes may be continued indefinitely to higher-orders, e.g. the next
stage would be to include
a dependence jointly upon $\bm,\bV,\bG$ in the PDF-{\em Ansatz}. All of the
results of this work carry
over to such closures in terms of higher-order time-derivatives, if one simply
considers the enlarged set
of moment-functions $\bhPsi=(\bhpsi,\bhV),(\bhpsi,\bhV,\bhG),$ etc. A linear
Langevin model will
always formally exist within such closure schemes which will exactly reproduce
the predictions of the
Rayleigh-Ritz effective action for the 2-time statistics of the
moment-variables $\bhPsi$ considered.

\vspace{.2in}

\noindent {\em (2.2) Rayleigh-Ritz Effective Action: Exact Expressions}

\noindent It is the main purpose of this paper to demonstrate the latter
essential fact. We will begin by developing
some exact expressions for $\Gamma_*[\bpsi]$. Substituting the given forms of
$\AAA(t),\PPP(t)$ into the action (\ref{2}),
one obtains
\be \Gamma= \sum_{i=1}^n \int_{t_0}^\infty
dt\,\,\alpha_i(t)[\dot{m}_i(t)-V_i(\bm(t),t)]. \lb{5} \ee
The overlap constraint (\ref{3}) may be incorporated by eliminating the
coefficient $\alpha_0(t),$ giving
\be \AAA(t) = 1 + \sum_{j=1}^n \alpha_j(t)(\hpsi_j(t)-m_j(t)) \lb{6} \ee
With that choice, the fixed expectation constraint (\ref{4}) becomes
\be \psi_i(t) = m_i(t) + \sum_{j=1}^n \alpha_j(t)C_{ij}(t) \lb{7} \ee
where $\bC(t):=\langle\bhpsi(t)\bhpsi^\top(t)\rangle_t-\bm(t)\bm^\top(t)$
defines the covariance matrix of the
moment-functions. Equation (\ref{7}) is easy to invert, with the result that
$\balpha(t)=\bC^{-1}(t)[\bpsi(t)-\bm(t)]$.
It is convenient to denote the inverse of the covariance matrix by
$\bGamma(t):=\bC^{-1}(t)$. Substituting the
above results for $\balpha(t)$, the action becomes
\be \Gamma_*[\bpsi;\bm]= \int_{t_0}^\infty
dt\,\,[\dot{\bm}(t)-\bV(\bm(t),t)]^\top \bGamma(t)
                         [\bpsi(t)-\bm(t)]. \lb{8} \ee
In this expression, all of the constraints have been properly incorporated, and
the only remaining
variational parameters are the $\bm(t)$-variables. A set of variational
equations must be developed
to determine these, derived from the stationarity condition of the action.

Before carrying out this variation, however, it is useful to introduce some
auxilliary quantities.
Define, within the PDF {\em Ansatz} employed, a single-time cumulant-generating
function
\be F(\bh|\bm(t),t):= \log\langle \exp(\bh\bdot\bhpsi(t))\rangle_t \lb{9} \ee
for the moment-functions $\bhpsi(t)$, where $\bm(t)$ is the mean of the
moment-function within the PDF {\em Ansatz}.
That is, the partial derivatives
\be C_{i_1\cdots i_p}^{(p)}(t)=\left. {{\partial^p F}
                     \over{\partial h_{i_1}\cdots \partial
h_{i_p}}}(\bh|\bm(t),t) \right|_{\bh=\bzed} \lb{10} \ee
are just the $p$th-order cumulants of the $\bhpsi(t)$ within the {\em Ansatz}.
The Legendre transform
\be H(\bmu|\bm(t),t): = \sup_{\bh}[\bmu\bdot\bh- F(\bh|\bm(t),t)], \lb{11} \ee
a {\em generalized entropy}, is the generating function of irreducible
correlation functions. That is,
\be \Gamma_{i_1\cdots i_p}^{(p)}(t)= \left. {{\partial^p H}
                     \over{\partial \mu_{i_1}\cdots \partial
\mu_{i_p}}}(\bmu|\bm(t),t) \right|_{\bmu=\bm(t)},  \lb{12} \ee
In particular, the relations hold that
\be \Gamma_{i}^{(1)}(t)= h_i(\bm(t),t),\,\,\,\,\,
    \Gamma_{ij}^{(2)}(t)= C^{-1}_{ij}(t)= {{\partial h_i}\over{\partial
\mu_j}}(\bm(t),t). \lb{13} \ee
The latter relation will prove crucial in what follows.

It may be worthwhile to explain the intuitive significance of these single-time
quantities before continuing with
the development of the formulae for the action. They are all part of a general
{\em thermodynamics of moments}.
Thus, the entropy $H$ is a form of Boltzmann's entropy, with his original sign
convention, i.e. positive and convex.
It is related to fluctuation probabilities of the empirical ensemble averages
$\overline{\bpsi}_N(t):={{1}\over{N}}
\sum_{n=1}^N \bhpsi^{(n)}(t)$ at time $t$ by
\be {\cal
P}\left(\left.\overline{\bpsi}_N(t)\approx\bmu\right|\bm(t),t\right)\sim
e^{-N\cdot H(\bmu|\bm(t),t)}, \lb{13a} \ee
where the samples $\bhpsi^{(n)}(t)$ are all independently chosen from the
ensemble $\PPP(\cdot|\bm(t),t)$. In other
words, $H\propto -\log {\rm Prob}$, which is Boltzmann's famous relation.
Because we have defined probabilities
with respect to the measure $\PPP(\cdot|\bm(t),t)$, this quantity corresponds
to what is in mathematics called
the {\em relative entropy}. The latter is an entropy of probability measures
analogous to Gibbs', but with
respect to an arbitrary {\em a priori} measure ${\cal P}$. Thus,
\be {\cal H}(\QQQ|\PPP):= \int \QQQ \log\left({{\QQQ}\over{\PPP}}\right).
\lb{13b} \ee
It is not hard to show that
\be H(\bmu|\bm(t),t)= \min_{\QQQ: \langle\bpsi(t)\rangle_\QQQ=\bmu}{\cal
H}\left(\QQQ|\PPP(\cdot|\bm(t),t)\right).
\lb{13c} \ee
This is a basic relation between the ``thermodynamic'' entropy
$H(\cdot|\bm(t),t)$ and the ``statistical mechanics''
measure $\PPP(\cdot|\bm(t),t)$. It is a form of the {\em maximum entropy
principle} \footnote{This a indeed
a maximum principle in terms of the usual entropies
$S(\bmu|\bm(t),t)=-H(\bmu|\bm(t),t)$, and ${\cal S}(\QQQ|\PPP)=
-{\cal H}(\QQQ|\PPP).$}. A good mathematical reference is \cite{Ellis}. As we
shall see later, $H_*(t):=
H(\bm(t)|\bm_*(t),t)$ should satisfy the second law of thermodynamics,
$dH_*(t)/dt<0$, when $\bm_*(t)$ is the predicted mean
history of the moment functions and $\bm(t)$ is any other solution of the
closure equations sufficiently near $\bm_*(t)$.
The derivatives of $H$ also have thermodynamic significance. For example, the
first derivatives $\bh(\bmu|\bm_*(t),t):=
{{\partial H}\over{\partial \bmu}}(\bmu|\bm_*(t),t)$ are the {\em thermodynamic
forces} which give the departure of the
moments $\bmu$ from the predicted means $\bm_*(t)$. Note, therefore, that
$\bh(\bmu|\bm_*(t),t)=\bzed$ if and only if
$\bmu=\bm_*(t)$. On the other hand, the Legendre transform
\be F(\bh|\bm(t),t): = \sup_{\bmu}[\bmu\bdot\bh- H(\bmu|\bm(t),t)], \lb{13d}
\ee
is a {\em generalized free energy}. It was defined already in (\ref{9}) above
via the logarithm of the
``partition function'' $Z(\bh|\bm(t),t):=\langle
\exp(\bh\bdot\bhpsi(t))\rangle_{\bm(t),t}.$

With this background, let us return to our analysis of the Rayleigh-Ritz
effective action. Equation (\ref{13}) yields
immediately a useful expression, complementary to (\ref{8}):
\be \Gamma_*[\bpsi;\bm]= \int_{t_0}^\infty
dt\,\,\left[{{d}\over{dt}}\bh(\bm(t),t)-
          {{\partial\bh}\over{\partial t}}(\bm(t),t)-
\bW(\bm(t),t)\right]^\top[\bpsi(t)-\bm(t)], \lb{14} \ee
where we have defined the new vector by matrix multiplication:
\be \bW(\bm(t),t): = \bGamma(\bm(t), t)\bV(\bm(t),t) . \lb{15} \ee
Indeed, it follows from the chain rule and (\ref{13}) that
\be {{d}\over{dt}}\bh(t)= \bGamma(t)\dot{\bm}(t)+{{\partial\bh}\over{\partial
t}}(t). \lb{16} \ee
Because of symmetry of $\bGamma,$ it follows that
$(\bGamma(t)\dot{\bm}(t))^\top=\dot{\bm}^\top(t)\bGamma(t)$.
Thus, we may use the previous relation to write
\be [\dot{\bm}(t)-\bV(t)]^\top\bGamma(t)
         = \left[{{d}\over{dt}}\bh(t)-{{\partial\bh}\over{\partial
t}}(t)-\bW(t)\right]^\top. \lb{17} \ee
When substituted into (\ref{8}), the result is (\ref{14}). This is a more
convenient form for variation. Indeed, setting
${{\delta\Gamma}\over{\delta m_k(t)}}=0$ gives
\be \sum_j\Gamma_{jk}[\dot{\psi}_j-\dot{m}_j]
    +\sum_j \left({{\partial\Gamma_{jk}}\over{\partial t}}+{{\partial
W_j}\over{\partial m_k}}\right)(\psi_j-m_j)
    +\sum_j [\dot{m}_j-V_j]\Gamma_{jk}=0, \lb{18} \ee
where the relations (\ref{13}),(\ref{17}) have again been employed.
Simplifying, we obtain finally
\be (\dot{\bpsi}-\bV)^\top\bGamma+ (\bpsi-\bm)^\top
        \left({{\partial\bGamma}\over{\partial t}}+{{\partial
\bW}\over{\partial \bm}}\right)=\bzed. \lb{19} \ee
This is the {\em variational equation} to determine $\bm(t)$ for a given
$\bpsi(t)$. When it is employed to eliminate
$\bm(t)$ in (\ref{8}) or (\ref{14}), the result is $\Gamma_*[\bpsi]$, the final
Rayleigh-Ritz approximation to the
effective action of the moment-variables $\bhpsi(t)$ in the closure.

One more transformation of the action is useful. We may write (\ref{8}) as the
sum of two terms:
\be \Gamma_*[\bpsi]= -\int_{t_0}^\infty
dt\,\,[\dot{\bpsi}-\dot{\bm}]^\top\bGamma[\bpsi-\bm]
                     +\int_{t_0}^\infty
dt\,\,[\dot{\bpsi}-\bV]^\top\bGamma[\bpsi-\bm]. \lb{20} \ee
In the first term we integrate once by parts, while in the second we use
(\ref{19}). This yields:
\be \Gamma_*[\bpsi]={{1}\over{2}}\int_{t_0}^\infty dt\,\,(\bpsi-\bm)^\top
\left[{{d}\over{dt}}\bGamma-2{{\partial\bGamma}\over{\partial t}}-
                {{\partial\bW}\over{\partial
\bm}}-\left({{\partial\bW}\over{\partial \bm}}\right)^\top\right]
                                  (\bpsi-\bm). \lb{21} \ee
Up until this point, no approximation has been made except Rayleigh-Ritz.
Equation (\ref{21}) is the most convenient
form to calculate the quadratic part of the Rayleigh-Ritz action.

\newpage


\noindent {\em (2.3) Quadratic-Order Action and Linear Langevin Model}

\noindent We shall now calculate the quadratic part of the full Rayleigh-Ritz
effective action $\Gamma_*[\bpsi]$.
An important quantity which appears is the {\em linear stability operator}
about a solution $\bm(t)$
of the moment-closure equations $\dot{\bm}=\bV(\bm,t)$, that is,
\be \bA(t):= {{\partial \bV}\over{\partial \bm}}(\bm(t),t). \lb{22} \ee
The subscript $*$ shall be used hereafter to indicate that the substitution of
the particular solution $\bm_*(t)$
for given initial data $\bm_{0*}$ has been made: thus, $\bA_*(t):= {{\partial
\bV}\over{\partial \bm}}(\bm_*(t),t).$
It is easy to relate ${{\partial \bW}\over{\partial \bm}}$ to the linear
stability operator. In fact from the
definition of $\bW$ in (\ref{15}) it follows that
\be {{\partial W_i}\over{\partial m_j}} = \Gamma_{ijk}^{(3)}V_k +
\Gamma_{ik}A_{kj}, \lb{23} \ee
where $\Gamma_{ijk}^{(3)}(t)={{\partial\Gamma_{ij}}\over{\partial
m_k}}(\bm(t),t)$ denotes the single-time
3rd-order irreducible correlation function within the PDF {\em Ansatz}. Using
this function again,
\be {{d}\over{dt}}\Gamma_{ij}= {{\partial \Gamma_{ij}}\over{\partial t}} +
\sum_k \Gamma_{ijk}^{(3)}\dot{m}_k =
                               (\partial_t+\dot{\bm}\bdot\grad_\bm)\Gamma_{ij}.
\lb{24} \ee
Then, by means of (\ref{23}),(\ref{24}), we can see that
\be {{d}\over{dt}}\bGamma-2{{\partial\bGamma}\over{\partial t}}-
                {{\partial\bW}\over{\partial
\bm}}-\left({{\partial\bW}\over{\partial \bm}}\right)^\top
    = -{{\partial\bGamma}\over{\partial
t}}+(\dot{\bm}-2\bV)\bdot\grad_\bm\bGamma-\bGamma\bA-(\bGamma\bA)^\top. \lb{25}
\ee
Furthermore, using $\dot{\bm}_*=V(\bm_*,t),$ it follows that
\be -{{\partial\bGamma_*}\over{\partial
t}}+(\dot{\bm}_*-2\bV_*)\bdot\grad_\bm\bGamma_*
     =-{{\partial\bGamma_*}\over{\partial
t}}-(\bV_*\bdot\grad_\bm)\bGamma_*=-{{d}\over{dt}}\bGamma_*. \lb{26} \ee
The effective action to quadratic order in deviations
$\delta\bpsi(t):=\bpsi(t)-\bm_*(t)$ from the solution $\bm_*(t)$
of the moment-equation is then found to be
\be \Gamma_*^{(2)}[\delta\bpsi]={{1}\over{2}}\int_{t_0}^\infty dt\,\,
(\delta\bpsi-\delta\bm)^\top
\left[-{{d}\over{dt}}\bGamma_*-\bGamma_*\bA_*-(\bGamma_*\bA_*)^\top\right]
                                  (\delta\bpsi-\delta\bm). \lb{27} \ee

In this expression, the quantity $\delta\bm(t):=\bm(t)-\bm_*(t)$ is to be
determined in terms of $\delta\bpsi(t)$ from
the variational equation (\ref{19}) linearized about the solution
$\bpsi_*(t)=\bm_*(t)$. It is convenient to rewrite
(\ref{19}) as
\be \dot{\bpsi}-\bV(\bpsi,t) + [\bV(\bpsi,t)-\bV(\bm,t)] + \bC
        \left[{{\partial\bGamma}\over{\partial t}}+\left({{\partial
\bW}\over{\partial \bm}}\right)^\top\right]
        (\bpsi-\bm)=\bzed. \lb{28} \ee
Using again (\ref{23}), (\ref{24}), it is then straightforward to linearize
this equation, yielding the variational
equation for
$\delta\bm(t)$:
\be
(\delta\dot{\bpsi}-\bA_*\delta\bpsi)
-2\bQ_*\bGamma_*(\delta\bpsi-\delta\bm)=\bzed, \lb{29} \ee
where the definition has been introduced
\begin{eqnarray}
2\bQ_* & := & -\bC_*[\dot{\bGamma}_*
               +\bGamma_*\bA_*+\bA_*^\top\bGamma_*]\bC_* \cr
     \,&  = & \dot{\bC}_*-\bA_*\bC_*-\bC_*\bA_*^\top. \lb{30}
\end{eqnarray}
Note that $\bL_*:= -\bA_*\bC_*$ is the {\em Onsager matrix}, in terms of which
the linearized closure equation may be
written in force-flux form: $\delta\dot{\bpsi}= -\bL_*\delta\bh.$ Then
(\ref{30}) may be restated as
\be \bQ_* = {{1}\over{2}}\dot{\bC}_* + \bL^s_*, \lb{31} \ee
with $\bL^s_*:={{\bL_*+\bL_*^\top}\over{2}}$ denoting the symmetric part of the
matrix $\bL_*$.

One may now obtain a final form for $\Gamma_*^{(2)}[\delta\bpsi]$ by
eliminating $\delta\bm(t)$ from (\ref{27})
by means of (\ref{29}) and by using the definition (\ref{30}) of $\bQ_*$. One
obtains
\be \Gamma_*^{(2)}[\delta\bpsi]={{1}\over{4}}\int_{t_0}^\infty
dt\,\,(\delta\dot{\bpsi}-\bA_*\delta\bpsi)^\top
\bQ_*^{-1}(\delta\dot{\bpsi}-\bA_*\delta\bpsi). \lb{32} \ee
This is the final result. One observes that it has the form of an {\em
Onsager-Machlup action}. That is,
the Rayleigh-Ritz result for $\Gamma_*^{(2)}[\delta\bpsi]$ is formally
equivalent to the effective action
that would be obtained for the solution of a linear Langevin model
$\delta\bpsi_+(t)$ of the fluctuation variable
$\delta\bhpsi(t)$. To be precise, the model stochastic equation is
\be \delta\dot{\bpsi}_+ = \bA_*(t)\delta\bpsi_+ + \bq(t) \lb{33} \ee
where $\bq(t)$ is a random force, white-noise in time, with zero mean and
covariance
\be \langle\bq(t)\bq^\top(t')\rangle = 2\bQ_*(t)\delta(t-t'). \lb{34} \ee
Note, in this context, that equation (\ref{31}) is the time-dependent
generalization of the {\em fluctuation-dissipation
relation} (of the 1st type), connecting the noise covariance matrix $\bQ_*$ and
the symmetric (dissipative) part
$\bL_*^s$ of the Onsager matrix. For details, see Section 6.

To understand the significance of the linear Langevin model, we must recall
some basic facts about the effective
action itself. As noted earlier, the effective action is a generating
functional for irreducible multi-time correlation
functions. That is, the $k$th-order irreducible correlator is given by
\be \Gamma^{(k)}_{i_1\cdots i_k}(t_1,...,t_k)
     = {{\delta^k
\Gamma}\over{\delta\psi_{i_1}(t_1)\cdots\delta\psi_{i_k}(t_k)}}[\bm_*].
\lb{34a} \ee
In particular, these coincide with functional Taylor coefficients in the series
expansion (\ref{g}). Furthermore,
the irreducible correlators of order up to $k$ determine all of the {\em
cumulants}---or connected correlators---
$C^{(k)}_{i_1\cdots i_k}(t_1,...,t_k)$ up to the same order $k$. \footnote{For
example, for $k=2$, $C_{i_1 i_2}(t_1,t_2)
=(\Gamma^{-1})_{i_1 i_2}(t_1,t_2)$; for $k=3$,
$C_{i_1i_2i_3}(t_1,t_2,t_3)=\sum_{j_1j_2j_3}\int ds_1\int ds_2\int ds_3\,\,
C_{i_1j_1}(t_1,s_1)C_{i_2j_2}(t_2,s_2)C_{i_3j_3}(t_3,s_3)
\Gamma_{j_1j_2j_3}(s_1,s_2,s_3)$; etc. See \cite{IZ}.} From these
two facts we see that knowledge of the Taylor series of $\Gamma_*[\psi]$ up to
terms of degree $k$ is equivalent to
knowledge of the Rayleigh-Ritz predictions for all multi-time correlators up to
order $k$. In particular, knowledge of the
quadratic term in the effective action, $\Gamma_*^{(2)}[\delta\bpsi]$, is
equivalent to knowledge of all two-time
correlators as predicted by Rayleigh-Ritz. Because knowledge of the linear
model is equivalent to knowledge of that
quadratic ``Onsager-Machlup'' part, the key conclusion that we draw is that
{\em the linear Langevin model is the unique
such model to reproduce exactly all the two-time correlators predicted by
Rayleigh-Ritz.} However, for correlators of
higher than second order, the two will in general disagree.

A simple observation which underlines this last point is the following: the
solution $\delta\bpsi_+(t)$ of the
linear Langevin model is always a {\em Gaussian} random function, while the
{\em true} fluctuation variable
$\delta\bhpsi(t)$ is in general {\em non-Gaussian}. Thus, although higher-order
cumulants than second are zero for the
Langevin solution $\delta\bpsi_+(t)$, they are generally non-zero for the true
fluctuation variable
$\delta\bhpsi(t)$. Of course, it is clear that for any random process
$\delta\bhpsi(t)$ there is a Gaussian
random process $\delta\bpsi_+(t)$ which has the same mean and variance (when
those exist). In fact, there is only
one such Gaussian process, in the sense that its distribution on the path-space
of histories is uniquely determined.
This result is sometimes called the Khinchin-Cram\'{e}r theorem. One way to
construct such a Gaussian process is
via the Central Limit Theorem.

The connection of the linear Langevin model to the Central Limit Theorem is
quite deep.
In fact, the precise empirical significance of the linear Langevin model is
that its predictions should be valid
for the normalized sum variable:
\be \delta\bhpsi_N(t):= {{1}\over{{\sqrt{N}}}}\sum_{n=1}^N
\delta\bhpsi^{(n)}(t), \lb{34b} \ee
in the limit $N\rightarrow\infty$, where the sum is over $N$ independent,
identically distributed samples. To prove this
fact, recall that the effective action is a fluctuation potential for the
empirical average over $N$ independent samples,
in the sense of equation (\ref{34c}). Now we consider the probability of a
small fluctuation value differing from the
ensemble-mean by terms of order $O\left({{1}\over{{\sqrt{N}}}}\right)$. That
is, we consider fluctuations
\be \bpsi(t)= \bm_*(t)+ {{\delta\bpsi(t)}\over{{\sqrt{N}}}}, \lb{34d} \ee
for $\delta\bpsi(t)=O(1)$. Substituting (\ref{34d}) into (\ref{34c}) and
employing the functional Taylor expansion
(\ref{g}) of $\Gamma[\bpsi]$, it is then straightforward to show that
\be {\rm Prob}\left(\delta\bhpsi_N(t) \approx \delta\bpsi(t)\right)
                    \sim \exp\left(-\Gamma^{(2)}[\delta\bpsi] +
O\left({{1}\over{{\sqrt{N}}}}\right)\right). \lb{34e} \ee
In the limit as $N\rightarrow\infty$ we arrive at the stated result. It is
clear that the distribution of $\delta
\bhpsi_N(t)$ is Gaussian in the limit. \footnote{In fact, we have just repeated
above one of the standard proofs in the
literature of the Central Limit Theorem!} Furthermore, $\Gamma^{(2)}[\bpsi]$
acts as the Onsager-Machlup action
of the limiting Gaussian variable, hence described also by the equivalent
linear Langevin model.

It is extremely important to emphasize that the existence of such a linear
Langevin model has only been
formally established, and only for the Rayleigh-Ritz approximation
$\Gamma_*[\bpsi]$. In general, the Central
Limit Theorem only guarantees that a Gaussian process should exist with the
same mean and covariance and not necessarily
a process obtained from a stochastic differential equation or, for that matter,
even a Markov process. It is thus
a very striking prediction of the Rayleigh-Ritz method with a ``Markovian''
{\em Ansatz}---and far from obviously true
---that the 2-time correlations should be reproducible by such a linear
Langevin model. Indeed, this prediction can fail
in a very striking way: the noise covariance $\bQ_*(t)$ given formally by
(\ref{30}) may turn not to be nonnegative!
Of course, nonnegativity is a fundamental requirement for any true covariance
function. If it fails, then the ``linear
Langevin model'' exists only in some formal sense and there is no actual
stochastic process which realizes the model.
Put another way, the Rayleigh-Ritz approximation $\Gamma_*[\bpsi]$ might fail
to satisfy the {\em realizability properties}
requisite for any true effective action. The relevant realizability properties
(positivity, unicity of minimizer, convexity)
have been discussed at length elsewhere \cite{Ey96,EyAlex1}. It is easy to see
that these properties of $\Gamma_*[\bpsi]$
will hold, at least for $\bpsi(t)$ close to the mean history $\bm_*(t)$, if and
only if $\Gamma^{(2)}_*[\delta\bpsi]\geq 0$,
with strict inequality for all $\delta\bpsi(t)\neq 0$. Furthermore, examination
of the Onsager-Machlup action (\ref{32})
shows that realizability of $\Gamma^{(2)}_*[\delta\bpsi]$ holds if and only if
the formal noise covariance $\bQ_*(t)$
appearing in the Langevin model is positive.

\section{A Simple Example}

It is interesting to compare the linear Langevin model with the full nonlinear
Rayleigh-Ritz. In general,
this should allow one to assess the limitations of the Langevin model for any
given problem, in particular,
to assess quantitatively how large are the corrections to its predicted
Gaussian statistics. To illustrate
the comparison of the Langevin dynamics and full Rayleigh-Ritz, we will discuss
here very briefly a 3-mode model
already considered in \cite{EyAlex1,EyAlex2}. This is a simple ``one-step
cascade'' model of dissipative turbulent
dynamics, originally introduced by Lorenz in 1960 \cite{Lor}. The dynamics are
just the Euler equations of a top, but
stochastically driven and linearly damped. The three modes are
$\bx=(x_1,x_2,x_3)$, of which the first is the driven,
unstable mode, and the second two are stable, damped modes. More specifically,
the equations of motion are given by
\be \dot{x}_i = A_i x_j x_k -\nu_i x_i + f_i, \lb{34f} \ee
with $i,j,k$ a cyclic permutation of $1,2,3$. Note that $A_1+A_2+A_3=0$ for
conservation of energy by the nonlinear
terms, the damping constants are $\nu_i>0,\,\,i=1,2,3$ and the random driving
forces are zero mean with covariance
\be \langle f_i(t) f_j(t')\rangle = 2\kappa_i\delta(t-t'), \lb{34g} \ee
all $\kappa_i>0,\,\,i=1,2,3$. A ``chi-square'' PDF {\em Ansatz} was proposed
for this system by Bayly, which leads to
the quasinormal closure equations. For full details of the model and closure,
we refer to \cite{EyAlex1,EyAlex2}.
Here we will simply remind the reader that the basic moments in the quasinormal
closure are the three modal energies
$\widehat{E}_i={{1}\over{2}}x_i^2, \,\,i=1,2,3$ and the triple moment
$\widehat{T}=x_1x_2x_3$, which gives
the energy transfer out of the unstable driven mode and into the stable, damped
modes. In the notations of this paper,
the closure dynamics is given by $\dot{\bm}=\bV(\bm)$ with
\be \bm=\left(\begin{array}{c}
               E_1 \cr
               E_2 \cr
               E_3 \cr
               T
               \end{array} \right)  \,\,\,\,\&\,\,\,\,
               \bV(\bm)=\left(\begin{array}{c}
                               A_1 T-2\nu_1E_1 +\kappa_1 \cr
                               A_2 T-2\nu_2E_2 +\kappa_2 \cr
                               A_3 T-2\nu_3E_3 +\kappa_3 \cr

4(A_1E_2E_3+A_2E_1E_3+A_3E_1E_2)-(\nu_1+\nu_2+\nu_3)T
                              \end{array} \right). \lb{34h} \ee
Although very simple, this model and closure will illustrate several key
features of our method. In addition to the
key comparison of full Rayleigh-Ritz and linear Langevin model, it will allow
us to discuss some important issues
concerning realizability.

The parameters of the linear Langevin model arising from the chi-square closure
for the 3-mode system are easy to determine.
The general form of the model is given in Eqs.(\ref{33})(\ref{34}). The
dynamical matrix $\bA$ (here time-independent,
since we consider only the statistical steady-state) is given by the
linearization of the closure equation, $\bA=
{{\partial \bV}\over{\partial \bm}}$, or,
\be \bA=\left(\begin{array}{cccc}
              -2\nu_1 & 0 & 0 & A_1 \cr
               0 & -2\nu_2 & 0 & A_2 \cr
               0 & 0 & -2\nu_3 & A_3 \cr
               4(A_2E_3+A_3E_2) & 4(A_1E_3+A_3E_1) & 4(A_1E_2+A_2E_1) &
-(\nu_1+\nu_2+\nu_3)
              \end{array} \right). \lb{34i} \ee
To obtain the matrix $\bA_*$ appearing in the Langevin model for a specific
situation, the corresponding
moments $\bm_*$ satisfying the fixed-point condition $\bV(\bm_*)=\bzed$ must be
substituted. The noise
covariance $\bQ$ can be calculated as the symmetric part of the Onsager matrix
$\bL$, and the latter
is derivable from $\bL= -\bA\bC$, once the matrix covariance $\bC$ of the
moment-functions $\bpsi$ is known.
The latter is provided by the PDF {\em Ansatz}, in this case Bayly's chi-square
{\em Ansatz}. A simple calculation
in that case gives
\be \bC=\left(\begin{array}{cccc}
               2E_1^2+{{3}\over{2}}T^{4/3} & {{1}\over{2}}T^{4/3} &
{{1}\over{2}}T^{4/3} &  2E_1T+3T^{5/3} \cr
               {{1}\over{2}}T^{4/3} & 2E_2^2+{{3}\over{2}}T^{4/3}&
{{1}\over{2}}T^{4/3} &  2E_2T+3T^{5/3} \cr
               {{1}\over{2}}T^{4/3} & {{1}\over{2}}T^{4/3} &
2E_3^2+{{3}\over{2}}T^{4/3} &  2E_3T+3T^{5/3} \cr
               2E_1T+3T^{5/3} & 2E_2T+3T^{5/3} & 2E_3T+3T^{5/3} &
\begin{array}{l}
                                                 8E_1E_2E_3+ \cr

4(E_1+E_2+E_3)T^{4/3}+19T^2
                                                                   \end{array}
              \end{array} \right). \lb{34j} \ee
Again, the matrices $\bC_*$ and, thence, $\bQ_*$ are obtained by substituting
the fixed-point moment values $\bm_*$.
It is worth emphasizing that $\bV$ and $\bC$ are the {\em only} statistical
inputs required from the PDF
{\em Ansatz} at the level of the linear Langevin model. If one is not
interested to carry out a full nonlinear
Rayleigh-Ritz calculation, then these are the only quantities that need be
provided {\em a priori} to construct
the linear model.

For the steady-state dissipative cascade of the 3-mode dynamics it is quite
easy to calculate both $\bA_*$
and $\bQ_*$. We have done so numerically with the same choice of parameter
values of the 3-mode model as in
our earlier work \cite{EyAlex1,EyAlex2}. The results given to four decimal
places are
\be \bA_*=\left(\begin{array}{cccc}
               -0.002 & 0.000 & 0.000 & 2.000 \cr
               0.000 & -2.000 & 0.000 & -1.000 \cr
               0.000 & 0.000 & -2.000 & -1.000 \cr
               -2.001 & -0.996 & -0.996 & -2.001
              \end{array} \right) \lb{34k} \ee
and
\be \bQ_*=\left(\begin{array}{cccc}
               3.385 & 0.545 & 0.545 & -6.868 \cr
               0.545 & 0.246 & -0.796 & 1.815 \cr
               0.545 & -0.796 & 0.246 & 1.815 \cr
               -6.868 & 1.815 & 1.815 & 8.428
              \end{array} \right). \lb{34l} \ee
The matrix $\bA_*$ can be easily checked to have all eigenvalues with negative
real parts. This indicates that
the closure fixed point $\bm_*$ is linearly stable. However, it turns out that
the putative noise covariance
$\bQ_*$ has eigenvalue spectrum $13.426,1.042,1.031$ and $-3.194$. One of the
eigenvalues is negative! Thus,
there is a breakdown of realizability in the Langevin model for this
dissipative cascade state.

Such a breakdown is also known to occur frequently in applications of the POP
method, which we shall discuss at some
length in Section 5. For example, Penland in her fundamental work \cite{Pen}
obtained negative eigenvalues for $\bQ_*$
in a POP analysis of a different quadratically-nonlinear 3-mode system, the
chaotic Lorenz model. Her interpretation
of this realizability breakdown is that it was due to nonlinearities of the
Lorenz model that could not be modelled
as white-noise random forces. This may be true, but it is not necessarily an
indication that the linear Langevin
model fails completely, for all statistics of the system. In our earlier work
\cite{EyAlex1,EyAlex2} we have
pointed out that the chi-square {\em Ansatz} for our 3-mode system---despite
its leading to a nonrealizable
linear Langevin model---nevertheless produces very good quantitative
predictions for several statistics. For example,
$E_2,E_3$ and $T$ are all predicted to within about $0.3\%$ and only the value
of $E_1$ is badly underpredicted
(by a factor of 3). Thus, simply labeling the model as ``bad'' because it leads
to a nonpositive noise covariance
$\bQ_*$ would be counterproductive, for good predictions would then be thrown
out with the bad ones. What is needed
are realizability diagnostics which are more focused and selective, that can
help to pinpoint precisely which
predictions are good and which bad.

In \cite{EyAlex1,EyAlex2} we have proposed that such diagnostics in the
statistical steady-state are provided by the
{\em effective potentials}. For any dynamical variable $\bhZ(t)$ of the system,
the effective potential $V(\bz)$
is a fluctuation potential for the empirical time-average $\overline{{\bf
Z}}_\rT:={{1}\over{\rT}}\int_0^\rT\,dt\,\bhZ(t)$.
That is,
\be {\rm Prob}\left(\overline{{\bf Z}}_\rT\approx \bz\right)\sim \exp(-\rT\cdot
V(\bz)), \lb{34m} \ee
in the limit as $\rT\rightarrow\infty$. Because the effective potential is a
measure of likelihood of fluctuations
in the very time-average used empirically to define the mean statistics, it is
plausible that it should be quite
sensitive to the failure of the closure for individual variables. The effective
potential can be obtained analytically
via the time-extensive limit of the effective action
\be V(\bz):= \lim_{\rT\rightarrow\infty}{{1}\over{\rT}}\Gamma[\bz_\rT],
\lb{34n} \ee
in which
\be \bz_\rT(t): = \left\{ \begin{array}{ll}
                         \bz & \mbox{for $0<t<\rT$} \cr
                         0 & \mbox{otherwise}
                        \end{array}                  \right. \lb{34o} \ee
Thus, it is easy to adapt the Rayleigh-Ritz algorithm to calculate the
effective potentials. In \cite{EyAlex1,EyAlex2}
we have applied the full nonlinear Rayleigh-Ritz algorithm in the 3-mode system
using the chi-square {\em Ansatz}
to calculate the effective potentials of modal energies $E_1$ and $E_2$ and of
the triple moment $T$. It was
found there that the potentials $V_{E_2}$ and $V_T$ are positive and convex,
satisfying realizability, whereas
the potential $V_{E_1}$ was negative and convex, i.e. realizability-violating.
In this case, therefore, the effective
potentials were---as conjectured---successful in discriminating the good
predictions from the bad.

Here we wish to calculate these same effective potentials, but using just the
linear model rather than full
Rayleigh-Ritz. In general, a linear Langevin dynamics such as (\ref{e}) gives
easily the joint effective potential
of all variables $\bhpsi$ therein, via the time-extensive limit of the
Onsager-Machlup action (\ref{eee}).
Thus, from (\ref{32}) one obtains directly the quadratic term
\be V^{(2)}_*(\bpsi) = {{1}\over{4}} \bK_*:\delta\bpsi\delta\bpsi, \lb{34p} \ee
with $\delta\bpsi=\bpsi-\bm_*$ and $\bK_*:= \bA_*^\top\bQ_*^{-1}\bA_*.$ Here
the dynamical matrix $\bA_*$
and noise covariance $\bQ_*$ in the Langevin model are evaluated at the
steady-state values $\bm_*$ of
the moment-averages. The effective potential of any single one of the moment
variables can then be obtained
by {\em minimizing} over the others:
\be V^{(2)}_*(\psi_i) = \min_{\psi_j,j\neq i} V^{(2)}_*(\bpsi). \lb{34q} \ee
Since the joint effective potential (\ref{34p}) is a simple quadratic form,
this minimization is easy to carry out.
In fact, if $\bK_*^{ii}$ is the minor matrix obtained from $\bK_*$ by deleting
the $i$th row and column and
$\bk_*^i$ is the vector obtained by deleting the element $k_*^{ii}$ from the
$i$th column, then
\be V^{(2)}_*(\psi_i) = {{1}\over{4}} \kappa_*^{ii}\delta\psi_i^2, \lb{34r} \ee
with $\kappa_*^{ii}:=k_*^{ii}-(\bK_*^{ii})^{-1}:\bk_*^i\bk_*^i$. These last
formulae allow a direct computation
of the effective potentials of moment-variables from the parameters appearing
in the Langevin model.

In Figs.1-3 we have plotted the parabolic effective potentials $V_{E_1},
V_{E_2}$ and $V_T$ obtained in this manner
from the linear Langevin model corresponding to the chi-square {\em Ansatz}.
The plots cover exactly the same range
as those in \cite{EyAlex1,EyAlex2}, where the potentials were calculated by the
full nonlinear Rayleigh-Ritz algorithm.
For comparison, we have plotted both pairs of potentials together in Figs.1-3,
the new ones using the linear Langevin model
and and the earlier ones from full Rayleigh-Ritz. Two points deserve to be
emphasized. First, the computational expense
of the Langevin model calculation is considerably lower than the full
Rayleigh-Ritz calculation. Each of the separate
symbols on the Rayleigh-Ritz effective potential curves was obtained by solving
numerically a fixed point problem, coming
from a perturbed closure equation. On the other hand, the Langevin model
calculation required the solution of {\em just one}
fixed point problem, to determine the mean moment values at the bottom of the
potentials. Those are all that are
needed to calculate the curvatures $\kappa_*^{ii}$ and hence the quadratic
potential curves via (\ref{34r}).
Thus, the number of fixed point symbols appearing in each of the Rayleigh-Ritz
curves is a quantitative
measure of the numerical superiority of the Langevin model calculation. Second,
we see that the
two calculation schemes lead to essentially equivalent results in this example,
at least for fluctuations
up to 20\% of the mean value. At least in this range, essentially the same
predictions for fluctuations
are obtained for the linear model as for full Rayleigh-Ritz, and at greatly
reduced expense.

Of course, over a wider range of fluctuations one should no longer expect that
the two calculation schemes will
agree. In general, the full Rayleigh-Ritz calculation should capture important
non-Gaussian fluctuation effects
that are missed by the simpler Langevin model. In Figs.4-5 are plotted over
wider ranges the two realizable
potentials in the 3-mode example, $V_{E_2}$ and $V_{T}$, calculated again both
by the full Rayleigh-Ritz method
and by the linear Langevin model. Clearly, for fluctuations 1-2 times the
means, the full Rayleigh-Ritz calculation
yields non-parabolic potentials associated to non-Gaussian statistics. The
range where the two calculations
agree gives an {\em a priori} indication of the size of fluctuations for which
the linear model may be
trusted. In the case of $V_{E_2}$ we see that fluctuations $\sim40\%$ of the
mean are well-described by
the linear model, while for $V_T$ the percentage is $\sim 60\%$. Of course, it
is an important question, not
just whether the Rayleigh-Ritz calculation gives different results, but whether
it {\em improves} upon the
predictions of the linear model. In \cite{EyAlex1,EyAlex2} it was already shown
that the Rayleigh-Ritz
effective potentials $V_{E_1}$ and $V_T$ give quite good quantitative results
for fluctuations over the smaller
ranges plotted in Figs.2-3. However, there the full Rayleigh-Ritz and linear
Langevin models substantially agree.
It is difficult to get accurate results for effective potentials in the wider
ranges directly from numerical
simulation of the 3-mode dynamics, because of the increasing rarity of those
large fluctuation events. Thus, we will
not show directly an improved agreement of the full Rayleigh-Ritz calculation
for the effective potentials in the
wider ranges. Nevertheless, one important observation can be made. Because
$x_2^2>0$ in {\em every} realization,
there must be zero probability for events with $E_2<0$. For this reason, the
true effective potential $V_{E_2}$
must blow up, i.e. diverge to positive infinity, as the negative values of
$E_2$ are approached. However,
because the effective potential predicted by the linear Langevin model is a
simple parabola, it will intersect
the ordinate axis $E_2=0$ at some {\em finite} value of $V$. It will thus
predict some positive probability
of seeing negative values $E_2<0$ (as would be true, if the fluctuations
$\delta E_2$ were indeed Gaussian random
variables.) However, as can be seen from Fig.4 the full Rayleigh-Ritz result
for $V_{E_2}$ is rising faster than the
parabolic potential from the linear model as $E_2\downarrow 0$. This is the
correct tendency, as indicated above,
and represents a qualitative improvement of the full Rayleigh-Ritz calculation.
In general, one may expect that
the full Rayleigh-Ritz calculation will give a more refined result, because it
uses more information both
from the dynamics and from the PDF {\em Ansatz} than the Langevin model.

Nevertheless, it is plausible to believe---again quite in general---that the
quadratic part will be the term which
dominates the effective potential sufficiently close to the minimum. The only
way to violate this expectation
is to have $\kappa_*^{ii}=0$ and $V(\psi_i)=O\left(\delta\psi^3_i\right)$.
Barring such cases of accidental
degeneracy, one can see that the quadratic term $V_*^{(2)}$ will well
approximate the full $V_*$ sufficiently near
to the minimum. In that regime, the linear Langevin model shall account for the
main tendencies of the full
theory. In particular, the Rayleigh-Ritz effective potential will satisfy
necessary realizability conditions
in the vicinity of the mean, when the Langevin model itself is realizability.
Thus, realizability of the
Rayleigh-Ritz effective potential is, close to the mean history, equivalent to
the realizability for a linear
Langevin model. It is thus particularly important  to understand the physical
hypotheses underlying the validity
of such a model.

\section{A Physical Derivation of the Linear Langevin Equation}

We shall now explain how exactly the same linear Langevin model can be obtained
from a more physically transparent
argument. Indeed, we shall show that the previous result for the quadratic
order action can recovered from a single
physical hypothesis: It is a basic assumption of the PDF-based moment closure
methodology that, to characterize a
probability distribution in phase-space, it is enough to know the mean values
it assigns to the moment functions.
In that case, the distribution is assumed to be described with sufficient
accuracy by the PDF {\em Ansatz} which
yields the same mean values for those moment functions. Let
$\langle\cdot|\bhpsi(s),s<t\rangle$ denote the expectation
over the conditioned ensemble given the past history $\{\bhpsi(s),s<t\}$ of the
moment variables before time $t$.
{\em Our basic assumption is that the PDF} Ansatz {\em can be employed as well
as an approximation for such an ensemble
conditioned on the past values.} More specifically, we shall assume that the
approximation is valid that
\be \langle\bhV(t)|\bhpsi(s),s<t\rangle \approx
\langle\bhV(t)\rangle_{\bhpsi(t),t}
                := \bV(\bhpsi(t),t). \lb{35} \ee
This is just a mathematical restatement of the hypothesis. Indeed, the
conditioned ensemble yields the expected values
$\langle\bhpsi(t)|\bhpsi(s),s<t\rangle=\bhpsi(t)$ and $P(\cdot;\bhpsi(t),t)$ is
thus the choice of the PDF {\em Ansatz}
which matches those expected values. \footnote{Note that $P(\cdot;\bhpsi(t),t)$
really means $\left.P(\cdot;\bm,t)
\right|_{\bm=\bhpsi(t)}$, i.e. the average over phase-space with respect to
$P(\cdot;\bm,t)$ is always taken first,
and then, subsequently, the {\em random} variable $\bhpsi(t)$ is substituted
for $\bm$.} We shall employ this hypothesis
mainly for the regime of small fluctuations, where
\be \bV(\bhpsi(t),t)=\bV(\bm_*(t),t)+ \bA_*(t)\delta\bhpsi(t)
+O\left(\delta\bhpsi^2\right) \lb{36} \ee
We have set $\delta\bhpsi(t):=\bhpsi(t)-\bm_*(t)$, the fluctuation variable.

We now consider the consequences of our hypothesis for the dynamics of the
fluctuations. We may write,
without any approximation,
\be \partial_t\bhpsi(t)= \bhV(t) = \langle\bhV(t)|\bhpsi(s),s<t\rangle +
\bhq(t), \lb{37} \ee
where the above equation is simply an implicit definition of the quantity
$\bhq(t)$:
\be \bhq(t):= \bhV(t) - \langle\bhV(t)|\bhpsi(s),s<t\rangle. \lb{38} \ee
It follows directly from this definition that $\langle\bhq(t)\rangle=0$ and
that
\be \langle\bhpsi(s)\bhq^\top(t)\rangle=0 \lb{39} \ee
for all $s<t$. If we now invoke our hypothesis in (\ref{37}), then we see that
\be \partial_t\bhpsi(t) \approx V(\bhpsi(t),t)+ \bhq(t). \lb{40} \ee
In other words, within the approximation considered, the {\em random }
moment-functions $\bhpsi(t)$
satisfy the same closure equations as the mean values
$\bm(t)=\langle\bhpsi(t)\rangle_t$, but with an
additional stochastic noise $\bhq(t)$ which is decorrelated from earlier values
of the moment-functions.
This is very similiar to the {\em regression hypothesis} made by Onsager,
according to which fluctuations
should decay on average according the same macroscopic equation obeyed by the
means. It is clear that
it is exactly at this point in the heuristic derivation that a ``Markovian''
approximation has been made.
It was emphasized by Onsager and Machlup (\cite{OM}, p.1509) that the
regression hypothesis is, for a Gaussian
random process, actually equivalent to the Markov property.

In the regime of small fluctuations $\delta\bhpsi(t)$, we may derive a more
specific formulation. There,
to linear order accuracy, the equation following from the hypothesis is
\be \delta\dot{\bhpsi}(t) \approx \bA_*(t)\delta\bhpsi(t) + \bhq(t). \lb{41}
\ee
Because of the linear relation, it is clear that consistency requires the force
$\bhq(t)$ to be white-noise in time.
Indeed, the equation (\ref{41}) can be solved explicitly, as
\be \delta\bhpsi(t)= \bG_*(t,t_0) \delta\bhpsi(t_0)+ \int_{t_0}^t
\bG_*(t,r)\bhq(r)\,dr \lb{42} \ee
where we have introduced the (retarded) matrix Greens function
\be \bG_*(t,t_0):= {\rm Texp}\left[\int_{t_0}^t
\bA_*(r)\,dr\right]\theta(t-t_0). \lb{43} \ee
It then follows by substituting $\delta\bhpsi(s)$ from (\ref{42}) into
(\ref{39}) that
\be \int_{t_0}^s \bG_*(s,r)\langle\bhq(r)\bhq^\top(t)\rangle\,dr =0 \lb{44} \ee
for all $s$ less than $t$. Differentiating with respect to $s$ then gives
\be \langle\bhq(s)\bhq^\top(t)\rangle =0 \lb{45} \ee
for all $s<t$. Thus, we see that the force must be delta-correlated in time
\footnote{In principle
the matrix $\bQ_*(t)$ in (\ref{46}) could be a differential operator with
finite-degree polynomial dependence on
$\partial_t$. We make the simplest assumption, that $\bQ_*(t)$ is an ordinary
matrix function}:
\be \langle\bhq(s)\bhq^\top(t)\rangle = 2\bQ_*(t)\delta(s-t). \lb{46} \ee
The noise covariance function $\bQ_*(t)$ is uniquely determined if we assume,
consistent with our hypothesis, that
the  fluctuation covariance
$\bC(t):=\langle\delta\bhpsi(t)\delta\bhpsi^\top(t)\rangle_t$ is the same as
$\bC_*(t)$
given by the PDF {\em Ansatz}. In that case, the noise covariance is uniquely
obtained from the relation
\be 2\bQ_*=\dot{\bC}_*-\bA_*\bC_*-\bC_*\bA_*^\top. \lb{47} \ee
Needless to say, we have now arrived at exactly the same linear Langevin model
that we obtained before from
the Rayleigh-Ritz approximation with a ``Markovian'' {\em Ansatz}. The present
derivation should make clearer the
physical assumptions involved in that more formal derivation.

The quadratic Onsager-Machlup term in the Rayleigh-Ritz effective action will
dominate in the vicinity
of the mean history, barring degenerate cases where the quadratic term
vanishes. As seen earlier
the realizability of the effective action in that region will be essentially
equivalent to the realizability of the
linear Langevin model (\ref{33}). The latter property is really a consistency
check on the validity of the physical
hypotheses underlying the Langevin model, in particular the consistency of
employing the PDF {\em Ansatz}
for an ensemble conditioned on the past history. In general, this depends upon
the particular situation
considered. In particular, enough variables must be included in the
moment-closure that the Markovian
assumption inherent in the approximation is justifiable.

This is perhaps the proper place to remind the reader that if a vector Markov
process $\bpsi(t)$ is
divided into two subsets $(\overline{\bpsi}(t),\bpsi'(t)),$ then, in general,
the separate subprocesses
$\overline{\bpsi}(t)$ and $\bpsi'(t)$ will not be Markov. Thus, if the
Rayleigh-Ritz effective action is determined
not for the {\em complete} set of moment-variables $\bpsi(t)$ but instead only
for a subset $\overline{\bpsi}(t)$,
then it will not ordinarily have the Onsager-Machlup form. However, there are
special cases in which this is true.
For example, suppose that the PDF {\em Ansatz} is such that the two variable
sets are uncorrelated at equal times:
\be \langle \bpsi'(t)\overline{\bpsi}^\top(t)\rangle-\langle
\bpsi'(t)\rangle\langle\overline{\bpsi}^\top(t)\rangle
    = {\bf O}. \lb{48} \ee
Suppose also that the closure equation of the ignored set of moments $\bm'(t)$
is independent of the retained set
$\overline{\bm}(t)$, that is,
\be \dot{\bm}'(t)=\bV'(\bm'(t),t), \lb{49} \ee
where $\bV'$ is a function of $\bm'$ alone. This would be realistic for cases
like fluid turbulence with a
passive scalar contaminant. In that case, the exact velocity dynamics is
independent of the passive scalar.
In cases where these two conditions hold, the Rayleigh-Ritz action
$\Gamma_*[\overline{\bpsi}]$ would
still have a quadratic part of the Onsager-Machlup form. This is
straightforward to show by arguments like
those used before. Needless to say, the two conditions are quite restrictive.

\section{Relation to Principal Oscillation Pattern (POP) Analysis}

There is a very close relation of the foregoing theory with the {principal
oscillation pattern} (POP) analysis,
particularly as it was developed by C. Penland \cite{Pen}. In her approach, the
POP method is a procedure
to derive directly from the empirical time series data for a selected set of
variables the linear Langevin
dynamics whose stochastic solution has the same mean and covariance as those
empirically derived, if such a
Langevin model exists. Her method can be explained in terms of the equations
used above. Indeed, assuming the
validity of a Langevin equation such as (\ref{e}), it is easy to show that, for
any $t>t'$,
\be {{d}\over{dt}}\bC(t,t')=\bA(t)\bC(t,t'). \lb{50} \ee
where $\bC(t,t'):=\langle\bhpsi(t)\bhpsi^\top(t')\rangle$ is the {\em
empirical} 2-time covariance matrix. Thus, the
linear dynamical matrix $\bA(t)$ can be obtained as
\be \bA(t)=\left.{{d}\over{dt}}\bC(t,t')\right|_{t'=t}\bC^{-1}(t), \lb{51} \ee
where, as before, $\bC(t):=\bC(t,t)$. Once $\bA(t)$ is known, the FDT relation
analogous to (\ref{47}) can be
used to determine $\bQ(t)$ from $\bA(t)$ and $\bC(t)$. This is essentially the
procedure proposed by Penland to
deduce the Langevin model from the data (with appropriate changes having been
made to allow for the general case
of time-dependent statistics considered here).

Of course, such a Langevin model need not exist at all, as pointed out also by
Penland. There are some basic
consistency properties that must be satisfied, if this is to be possible.
First, the computed noise covariance
$\bQ(t)$ must be positive-definite. This is the same type of realizability
condition that was encountered in the
Rayleigh-Ritz approach. It is a qualitative check of the Langevin model
assumption, basically amounting to a
statistical stability condition within that framework. A more stringent and
quantitative property for validity
of a linear Langevin model is deduced by inverting (\ref{50}), to obtain
\be \bA(t)={{d}\over{dt}}\bC(t,t')\cdot\bC^{-1}(t,t'). \lb{52} \ee
This must hold for {\em all} $t'<t$. In general, the righthand side of
(\ref{52}) defines an object $\bA(t,t')$
for $t'=t-\tau<t$ which will have a nontrivial dependence upon the time lag
$\tau$. To be consistent with a
linear Langevin model, however, there should be no such dependence. Hence, the
degree of constancy of
$\bA(t,t')$ in the lag time $\tau$ is a quantitative measure of the validity of
the linear Langevin
modeling assumption. This is Penland's ``$\tau$-test'' \cite{Pen}.

Although the Rayleigh-Ritz and POP methods are seen to be closely related, they
have almost opposite points of view.
The Rayleigh-Ritz approach is an {\em a priori} theoretical method, whereas the
POP approach is {\em a posteriori}
and empirical. That is, the Rayleigh-Ritz method uses the underlying dynamical
equations of motion computationally,
in conjunction with physically-inspired guesses for the system statistics.
Thus, it deduces the linear Langevin model
without any direct empirical input (aside from experimental knowledge which may
have been exploited to develop suitable
PDF {\em Ans\"{a}tze} for the problem). On the other hand, the POP method makes
no use of the dynamical equation of
motion, and, indeed, could be applied to time series generated by very
different means than a dynamical equation.
POP is blind to theoretical considerations, except through the choice of
relevant variables $\bhpsi(t)$ to be
used in the analysis. Because the two approaches have such different
philosophies but yet a close formal relationship,
they should be quite complementary in assaulting a given problem. In both
cases, a linear Langevin model is obtained
which is supposed to reproduce faithfully all first- and second-order
correlators of the selected set of variables.
Thus, the Langevin model deduced theoretically by Rayleigh-Ritz may be compared
directly with that deduced from the
experimental data via POP. On the other hand, a successful application of the
empirical POP procedure for a given
set of variables---with ``success'' meaning here that realizability of the
noise covariance is satisfied and that
lag-dependence of the deduced matrix $\bA(t,t')$ is weak---would imply the
possibility of carrying out a successful
Rayleigh-Ritz approximation with the same set of variables.

Although POP is an {\em a posteriori} method, relying upon a substantial
empirical input, it has predictive power.
This capability rests upon a basic hypothesis: that the POP Langevin model,
while constructed only to reproduce
partially the second-order statistics, may also be used to predict other
statistical properties of the system
with some accuracy. In particular, quantities such as transition probability
densities $P(\bpsi,t|\bpsi_0,t_0)$ can be
deduced from the POP Langevin model. For many problems of weather and climate
prediction, such probabilities would yield
crucial information. For example, a measure of the spread of the predictions,
such as Penland's {\em relative
discrepancy}:
\be \delta(t,t_0):= {{\langle
\|\bhpsi(t)-\bG(t,t_0)\bhpsi(t_0)\|^2\rangle}
\over{\langle\|\bhpsi(t_0)\|^2\rangle}}, \lb{53} \ee
can be estimated. This quantity is itself second-order, but not one used in the
derivation of the POP Langevin
model and thus not one that the model is guaranteed to predict successfully for
arbitrary time lags $\tau=t-t_0$.

The Rayleigh-Ritz method has the potential for superior predictive ability,
particularly with regard to
non-Gaussian statistics and large fluctuations. As we have noted, the full
Rayleigh-Ritz calculation predicts
non-vanishing higher-order cumulants of the moment variables, as required for
non-Gaussian statistics. Thus, when the
statistics of the problem---such as the transition probabilities---have a very
non-Gaussian form, the Rayleigh-Ritz
approximation may still derive them successfully. Previous work on simple
systems has already shown that very large
fluctuations, far outside the Gaussian core, may be successfully captured by a
Rayleigh-Ritz calculation.
See the examples in \cite{EyAlex2}. Thus, the Rayleigh-Ritz method can yield
crucial information about such
large fluctuations, not available by a POP analysis. When the system is
strongly fluctuating, and the most
probable future event is only weakly selected, realizations deviating from that
predicted event by percentages $\gg
\delta(t,t_0)$ would have sizable probability. In that case, a Gaussian
transition density, such as always yielded
by a linear Langevin model, would yield very misleading estimates of event
probabilities. The Rayleigh-Ritz
method has the potential to predict better the non-Gaussian probabilities of
such large-deviation events.

\section{Fluctuation-Dissipation Relations}

A basic premise of our work is that the dynamical system considered is
statistically stable, i.e. that the
probability measures $\PPP(t)$ which solve the Liouville equation
$\partial_t\PPP(t)=\hL\PPP(t)$ for all
initial conditions $\PPP_0$ converge to a unique invariant measure
$\PPP_\infty$ as $t\rightarrow\infty$.
\footnote{This remark applies to autonomous evolution only, in which the
Liouville operator $\hL$ has no explicit
time-dependence.} Of course, such statistical stability is not
precluded---indeed, is even assisted---by chaotic
instability of the underlying microdynamics. In this context, one expects that
a generalized second law of
thermodynamics should apply, appropriate to dissipative dynamical systems that
are driven by external forces
or open to the environment. In such a circumstance the usual thermodynamic
entropy of the system proper obviously
need not increase, but only the overall entropy of the system plus environment.
An entropy function appropriate
to describe the irreversible decay of the system to its stable, dissipative
steady-state is provided by the
{\em relative entropy} introduced in Section 2. Its production rate (or,
rather, destruction rate, with our sign
convention) is zero in the steady-state itself, and thus does not account for
the dissipative processes occuring
therein. The latter have been subtracted out, in the definition of the relative
entropy. However, the relative
or generalized entropy turns out to be the most useful concept in the dynamical
description of the system
proper, since it provides a {\em Lyapunov functional} for the irreversible
decay to the statistical steady-state.
Furthermore, the usual relations between random fluctuations and mean
dissipation---the {\em fluctuation-dissipation
relations}---are valid in terms of this generalized or ``excess'' entropy
production within the Rayleigh-Ritz approximation,
subject to satisfaction of realizability constraints. Such results for
statistical steady-states, under hypotheses
paralleling those made here, are due originally to Schl\"{o}gl \cite{Schlog}.
In view of the generality of these results,
it is appropriate to give here a brief account.

The generalized entropy relative to the predicted mean history $\bm_*(t)$ is
defined by
\be H_*(\bmu,t):= H(\bmu|\bm_*(t),t), \lb{55} \ee
where $H$ on the righthand side is given by (\ref{11}) in the text. The
generalized entropy production or
excess dissipation is then defined by
$\eta_*(\bmu,t):=\left.{{d}\over{dt'}}H_*(\bm(t'),t')\right|_{t'=t}$ where
$\bm(\cdot)$ is the solution of the closure equation which satisfies
$\bm(t)=\bmu$. Thus, a simple calculation gives
\be \eta_*(\bmu,t)=\bh_*(\bmu,t)\bdot\bV(\bmu,t) +{{\partial H_*}\over{\partial
t}}(\bmu,t). \lb{56} \ee
It is not hard to show that, to quadratic order accuracy in small deviations,
\be \eta_*:= -\left(\bL^{s}_* +
{{1}\over{2}}\dot{\bC}_*\right):\delta\bh\delta\bh +O\left(\delta h^3\right).
\lb{57} \ee
We shall sketch the proof below. First, however, let us recall the relation
between the noise covariance $\bQ_*(t)$
and the Onsager matrix $\bL_*(t)= -\bA_*(t)\bC_*(t)$, already given in
(\ref{31}):
\be \bQ_*= \bL^{s}_* + {{1}\over{2}}\dot{\bC}_*. \lb{54} \ee
This is the {\em fluctuation-dissipation relation of the first type}. Along
with (\ref{57}) it allows one to
express the quadratic part of the entropy production (or dissipation) directly
in terms of the noise covariance $\bQ_*$:
\be \eta_*:= -\bQ_*:\delta\bh\delta\bh +O\left(\delta h^3\right). \lb{58} \ee
Thus, the FDR of 1st type expresses a direct connection between the noise
characteristics and the dissipative part of
the linear dynamics. The generalized entropy $H_*(\bmu,t)$ is a nonnegative,
convex function, vanishing at $\bm_*(t)$.
When the noise covariance $\bQ_*(t)$ is positive-definite, (\ref{58}) implies
that the generalized entropy is a Lyapunov
function for the closure dynamics. At least for small deviations $\delta\bm(t)$
from the solution $\bm_*(t)$, where
the linearized dynamics $\delta\dot{\bm}(t)= -\bL_*(t)\delta\bh(t)$ applies,
the entropy $H_*(t)$ is guaranteed
to decrease in time, $dH_*(t)/dt<0.$ This implies stability of the history
$\bm_*(t)$ under the closure dynamics,
generalizing the results of Schl\"{o}gl \cite{Schlog} to non-steady states of
strongly fluctuating systems.

The proof of (\ref{57}) is as follows: Expanding the generalized entropy in a
power series about $\bm_*(t)$ yields,
with $\delta\bmu(t):=\bmu-\bm_*(t)$,
\be H_*(\bmu,t)= {{1}\over{2}}\bGamma_*(t):\delta\bmu(t)\delta\bmu(t)
                 +
{{1}\over{3!}}\bGamma_*^{(3)}(t)\dot{:}\delta\bmu(t)\delta\bmu(t)\delta\bmu(t)
                 + O\left(\delta\mu^4\right). \lb{59} \ee
Recall that $\bGamma_*^{(p)}(t):={{\partial^p
H_*}\over{\partial\bmu^p}}(\bm_*(t),t).$  Taking one derivative
of (\ref{59}) with respect to $\bmu$ gives
\be \bh_*(\bmu,t) = \bGamma_*(t)\cdot\delta\bmu(t)+
{{1}\over{2}}\bGamma_*^{(3)}(t):\delta\bmu(t)\delta\bmu(t)
                 + O\left(\delta\mu^3\right). \lb{60} \ee
Introducing $\delta\bh(t):=\bGamma_*(t)\cdot\delta\bmu(t)$, the latter becomes
\be \bh_*(\bmu,t) = \delta\bh(t)+
{{1}\over{2}}\bGamma_*^{(3)}(t):\delta\bmu(t)\delta\bmu(t)
                 + O\left(\delta\mu^3\right). \lb{61} \ee
A similiar Taylor expansion of the dynamical vector field $\bV$ gives
\begin{eqnarray}
\bV(\bmu,t) & = & \bV_*(t) + \bA_*(t)\delta\bmu(t) 
                                    + O\left(\delta\mu^2\right) \cr
         \, & = & \bV_*(t) - \bL_*(t)\delta\bh(t) + O\left(\delta h^2\right),
\lb{62}
\end{eqnarray}
with $\bV_*(t):=\bV(\bm_*(t),t)$. Now the Taylor expansion of the first part of
the excess dissipation (the force-flux
quadratic form) can be obtained by direct substitution of
(\ref{61}),(\ref{62}):
\be \bh_*\bdot\bV= \delta\bh\bdot\bV_* - \bL_*:\delta\bh\delta\bh
                   +
{{1}\over{2}}\bGamma_*^{(3)}\dot{:}\delta\bmu\delta\bmu\bV_*+ O\left(\delta
h^3\right). \lb{63} \ee
The second part of the entropy production is obtained by partial
differentiation of (\ref{59}):
\be {{\partial H_*}\over{\partial t}}= -\delta\bh\bdot\bV_*
    +{{1}\over{2}}{{\partial \bGamma_*}\over{\partial
t}}(\bm_*(t),t):\delta\bmu\delta\bmu
    +{{1}\over{3!}}{{\partial \bGamma_*^{(3)}}\over{\partial
t}}(\bm_*(t),t)\dot{:}\delta\bmu\delta\bmu\delta\bmu
                 + O\left(\delta\mu^4\right). \lb{64} \ee
We made use of the facts that ${{\partial}\over{\partial t}}\delta\bmu(t) =
{{\partial}\over{\partial t}}(\bmu-\bm_*(t))
= -\bV_*(t)$ and that, for every non-negative integer $p$,
$\bGamma_*^{(p+1)}(\bm_*(t),t) ={{\partial \bGamma^{(p)}}
\over{\partial\bmu}}(\bm_*(t),t).$  Adding together the two parts of the
entropy production from (\ref{63}),(\ref{64})
then gives
\be \eta_*= -\bL_*:\delta\bh\delta\bh +
{{1}\over{2}}\left(\bGamma^{(3)}_*\bdot\bV_*
    + {{\partial\bGamma_*}\over{\partial t}}\right):\delta\bmu\delta\bmu
+O\left(\delta h^3\right). \lb{65} \ee
If one recalls that ${{d}\over{dt}}\bGamma_*=\bGamma^{(3)}_*\bdot\bV_* +
{{\partial\bGamma_*}\over{\partial t}}$,
as in (\ref{24}), then we obtain finally
\begin{eqnarray}
\eta_* & = &
-\bL_*:\delta\bh\delta\bh+{{1}\over{2}}\dot{\bGamma}_*:\delta\bmu\delta\bmu
+O\left(\delta h^3\right) \cr
        \,& = &
-\left(\bL_*^s+{{1}\over{2}}\dot{\bC}_*\right):\delta\bh\delta\bh+O\left(\delta
h^3\right), \lb{66}
\end{eqnarray}
where $\dot{\bGamma}_*= -\bGamma_*\dot{\bC}_*\bGamma_*$ was employed in the
last line. This is just (\ref{57}),
as was claimed.

There is another result, {\em the fluctuation-dissipation relation of the
second type}, which holds for a
general linear Langevin model. This relation expresses a proportionality
between the mean response function
to an appropriately coupled force and a time-derivative of the 2-time
correlation function. The equation
to be considered is
\be \delta\dot{\bpsi} = -\bL_*(t)(\delta\bh-\bff(t)) + \bq(t) \lb{67} \ee
where $\bff(t)$ is a deterministic external force. Because of the linearity of
this equation, it follows immediately
that the corresponding response function $\bH(t,t_0):={{\delta
\bpsi(t)}\over{\delta \bff(t_0)}}$ is {\em non-random}
and its average $\bH_*(t,t_0):=\langle\bH(t,t_0)\rangle$ is thus given just by
the solution of
\be {{\partial}\over{\partial t}}\bH_*(t,t_0) = \bA_*(t)\bH_*(t,t_0) +
\bL_*(t_0)\delta(t-t_0). \lb{68} \ee
It is not hard to see that the latter solution is
\be \bH_*(t,t_0)= \bG_*(t,t_0)\bL_*(t_0), \lb{69} \ee
with $\bG_*$ the matrix Greens function defined in (\ref{43}). On the other
hand, by using the same Greens function
to solve (\ref{50}) for $\bC_*(t,t_0)$ starting from time $t_0$, it is
determined that
\be \bC_*(t,t_0)= \bG_*(t,t_0)\bC_*(t_0). \lb{70} \ee
Because $\bL_*(t_0):=-\bA_*(t_0)\bC_*(t_0)$ and because
${{\partial}\over{\partial t_0}}\bG_*(t,t_0)
= -\bG_*(t,t_0)\bA_*(t_0)$ for $t>t_0$, it follows from (\ref{69}),(\ref{70})
that
\be {{\partial}\over{\partial t_0}}\left[\bC_*(t,t_0)\bGamma(t_0)\right] =
\bH_*(t,t_0)\bGamma(t_0) \lb{71} \ee
for $t>t_0$. This proportionality is termed an FDR of 2nd type. The reader
should be cautioned at this point that
there is a great divergence of terminology in the literature. One finds often
the following terms used instead:
{\em FDR of 1st kind} to indicate what we called FDR of 2nd type and {\em FDR
of 2nd kind} to indicate our FDR
of 1st type. There are also authors who call (\ref{70}) the FDR of 2nd type
(1st kind) rather than (\ref{71})!
For further discussion of these matters, see \cite{ELS}, Sections 3.2, 4.1 and
Appendix D.

Another relation of the ``2nd type'' exists within the Rayleigh-Ritz formalism.
This has a slightly different
character, in that the external perturbation field is now added to the {\em
deterministic} equation. As
the simplest example, consider the following perturbation of the moment-closure
equations:
\be \dot{\bm}=\bV(\bm,t) + \bC(\bm,t)\bdot\bh(t), \lb{72} \ee
in which $\bC(\bm,t)$ is the model single-time covariance matrix provided by
the PDF {\em Ansatz}.
Then, if $\bR(t,t_0):=\left.{{\delta \bm(t)}\over{\delta
\bh(t_0)}}\right|_{\bh=\bzed}$ is the corresponding
response function, it is easy to see by functional differentiation that, for
initial conditions $\bm(t_0)=\bm_{*0}$
in the above equation, $\bR_*$ satisfies
\be \partial_t \bR_*(t,t_0)= \bA_*(t)\bR_*(t,t_0) + \bC_*(t_0)\delta(t-t_0),
\lb{73} \ee
The solution is just $\bR_*(t,t_0)= \bG_*(t,t_0)\bC_*(t_0)$ for $t>t_0$. Thus,
we see by reference to (\ref{70}) above
and the symmetry of the covariance that
\be \bC_*(t,t_0)= \bR_*(t,t_0) + \bR_*^\top(t,t_0), \lb{74} \ee
where $\bR_*^\top(t,t'):= [\bR_*(t',t)]^\top.$ The relation (\ref{74}) might be
better termed a {\em fluctuation-response
relation}, in analogy to that of Kraichnan \cite{Kr59}. It turns out that this
relation is completely general within
the Rayleigh-Ritz method. In fact, equation (\ref{72}) above is nothing more
than the Euler-Lagrange equation
for $\bm(t)$ in the Rayleigh-Ritz algorithm, when the expectation constraint
(\ref{4}) is incorporated via a Lagrange
multiplier $\bh(t)$. That is, equation (\ref{72}) above is equivalent to
equation (3.93) in \cite{Ey96}. All of these
statements remain true even when the random variables whose 2-time covariance
is to be approximated by Rayleigh-Ritz
are not the basic moment-variables appearing in the closure and the linear
Langevin model is not available.
The demonstration of this fact will be given elsewhere \cite{Ey98II}, since it
is outside the scope of the present work.
The result (\ref{74}) is very useful, because it provides the most efficient
numerical procedure to extract the Rayleigh-Ritz
predictions for the 2-time covariances.

\section{Conclusions}

In this work we have shown how the general Rayleigh-Ritz algorithm for
statistical dynamics of nonlinear systems,
proposed in \cite{Ey96}, gives rise to linear Langevin models. Such stochastic
models reproduce the predictions of the
full Rayleigh-Ritz calculation for 2nd-order statistics. In general, for higher
order statistics and larger fluctuations,
the two methods yield different predictions. Thus, the Rayleigh-Ritz approach
gives also a means to assess {\em a priori}
the domain of validity of the linear Langevin models. A more physical
derivation of Langevin models was also sketched.
The theoretical and {\em a priori} Rayleigh-Ritz method was compared with the
empirical and {\em a posteriori}
POP method of Penland. Finally, some general results on the thermodynamics of
statistical moments---law of entropy
increase and fluctuation-dissipation relations at the linear level---were
derived within the Rayleigh-Ritz formalism.

\noindent {\bf Acknowledgements:} I wish to thank Frank Alexander for many
useful conversations on the subject of this
work and for informing me of the POP method of Penland. I also wish to thank
Shiyi Chen and CNLS for support to
attend the 18th Annual International Conference of CNLS on ``Predictability''.
My conversations with many of
the participants and their lectures helped considerably in sharpening the focus
of this work.

\end{document}